\documentclass[%
nofootinbib,
 reprint,
 amsmath,amssymb,
 aps,pra, superscriptaddress
]{revtex4-2}
\usepackage{graphicx}
\usepackage{amsmath}
\usepackage{nicefrac}
\usepackage[hidelinks]{hyperref}
\usepackage{braket}
\usepackage{amssymb}
\usepackage{siunitx}
\usepackage{xspace}
\usepackage{bm}
\usepackage{comment}
\usepackage{subcaption}
\newcommand{\TK}[1]{{#1}}
\begin{document}
\title{
{Laboratory Frame Representation of Time-Dependent Gravitational Waveforms
}
}
\author{Lars Fischer}
\altaffiliation[Present address: ]{Institute for Quantum Electronics \& Quantum Center, ETH Zurich, 8093 Zurich, Switzerland}
\affiliation{Universität Hamburg,\\
Luruper Chaussee 149, 22761 Hamburg, Germany}

\author{Tom Krokotsch}%
\email{Contact author: tom.krokotsch@desy.de}
\affiliation{Universität Hamburg,\\
Luruper Chaussee 149, 22761 Hamburg, Germany}

\author{Gudrid Moortgat-Pick}
\affiliation{Universität Hamburg,\\
Luruper Chaussee 149, 22761 Hamburg, Germany}
\affiliation{Deutsches Elektronen-Synchrotron DESY,\\
 Notkestraße 85, 22607 Hamburg, Germany}

\begin{abstract}
Next-generation gravitational wave (GW) experiments will explore higher frequencies where wavelengths become comparable to the dimensions of the detector.
In this regime, GWs may be detected not only through mechanical deformations from tidal forces, but also via induced effective currents in electromagnetic background fields. 
However, the calculation of these signals often requires transforming the GW metric to all orders into the laboratory frame of the detector.
Here, we derive a closed-form expression for the metric transformation of general chirp-like waveforms expressed in terms of the transverse-traceless GW metric components, their integral, and their derivative. 
For more complex signals, where analytical integration is impractical, we provide an efficient approximation based on Taylor expansions of the retarded time to coalescence. 
\TK{We find that quasi-Newtonian chirp signals can be accurately transformed using a monochromatic approximation, whilst also demonstrating that more complex waveforms require the general transformation which we derive.}
Thus, the calculated transformations provide essential tools for designing high-frequency GW experiments and analyzing signals from compact object mergers at MHz to GHz frequencies.
\end{abstract}

\maketitle

%
%
\section{Introduction}
Einstein’s field equations in general relativity (GR) are covariant under coordinate transformations, which can often be solved in coordinate systems that are mathematically convenient but often physically unintuitive.
Thus, ever since gravitational waves (GWs) were first proposed, their physical effects have been continuously discussed \cite{EINSTEIN193743, BONDI1957, rakhmanov_fermi-normal_2014}. For detector development and signal processing, it is essential to predict the response of a detector to a passing GW to high accuracy. 
In particular, detector structures like stationary masses or static magnetic fields can usually only be established in the local laboratory frame of the detector, where coordinates are assumed to be rigid in nearly flat spacetime. This requires the construction of Fermi normal coordinates, where the metric components $g_{\mu\nu}(x)$ obey $g_{\mu\nu}(0)=\eta_{\mu\nu}$, with $\eta_{\mu\nu}$ the Minkowski metric, and $\partial_\rho g_{\mu\nu}(0)=0$ when choosing the origin $x_0=0$ to be the center of mass of the detector. 
Locally, this resembles an inertial frame of the detector, often referred to as the proper detector (PD) frame \cite{rakhmanov_fermi-normal_2014}.

In contrast, wave solutions are most easily obtained from the linearized Einstein equations in the transverse-traceless (TT) gauge, where coordinate distances between freely falling masses are not affected by GWs. For experiments based on Michelson interferometers, the gauge invariant travel time of the light can be calculated entirely in the TT frame \cite{maggiore_gravitational_2008}. 
However, this approach is not applicable when detector geometries are more complex \cite{berlin_detecting_2023, domcke_novel_2022}, or the detector resonantly responds to the GW signal \cite{Weber_Bars}. 
For experiments of this kind, the transformation of GWs into PD coordinates is the preferred method to model the detector response in terms of tidal forces or effective electromagnetic currents \cite{Weber_GW_detection}. Closed analytical expressions for exact transformations between a linearized TT and PD metric have already been derived in \cite{Fortini1982} and were generalized in \cite{marzlin_fermi_1994} using infinite sums in growing orders of $\omega_gL_\text{det}$, where $L_\text{det}$ is the detector size and $\omega_g=2\pi f_g$ the GW frequency. 
This usually allows considering a long-wavelength limit $\omega_gL_\text{det}\ll 1$ where only the first terms of the sum need to be considered \cite{berlin_mago20_2023, pappas2025highfrequencygravitationalwavesearch}.
However, a growing interest in high-frequency GW searches, where $\omega_g>\text{kHz}$, \cite{aggarwal_challenges_2025} necessitates calculations where the GW wavelength is of the same order as the detector size. In this regime, all orders of the transformation in \cite{Fortini1982, marzlin_fermi_1994} must be used, which becomes computationally challenging.

Recently, \cite{berlin_detecting_2023} provided an analytic simplification of the transformation for monochromatic plane GWs and applied it to calculate geometric overlaps between effective GW currents and microwave cavity modes in static magnetic fields. However, most astrophysical GW sources produce chirp-like signals with time-dependent frequency and amplitude.

To address this limitation, we extend this analytic framework to practically transform arbitrary gravitational waveforms from TT coordinates into the PD frame at all orders. We focus in particular on chirp signals generated by inspiraling compact binaries. When an exact integral evaluation in the exact form proves challenging, we provide efficient approximations. 

The importance of such analytic extensions becomes clear when considering practical GW detection. In GW searches, analytic formulas are crucial because they enable efficient signal correlation with theoretical waveform templates, which can significantly improve detection sensitivity \cite{maggiore_gravitational_2008}. Given that numerous templates must be scanned during data analysis, closed-form expressions are essential for computational efficiency.

In signal analysis, chirp waveforms are particularly interesting, as they represent the only class of coherent GWs detected so far. They can be generated by coalescence of two compact objects like black holes or neutron stars. Similar signal forms are also expected to be produced by various high-frequency sources such as mergers of primordial black holes \cite{franciolini_hunt_2022} or exotic dense objects like Bose stars \cite{chung-jukko_multimessenger_2024}.
Our results improve modeling the response of detectors to such high-frequency GWs, which includes microwave cavities in static magnetic fields \cite{berlin_detecting_2023, schmieden_global_2023, reina-valero_high-frequency_2025}, lumped-element detectors in static magnetic fields \cite{domcke_novel_2022} and microwave cavities loaded with RF power \cite{ballantini_microwave_2005, berlin_mago20_2023, fischer2024characterisationmagocavitysuperconducting}. 
Electromagnetic axion detectors also operate at $\omega_gL_\text{det}\gtrsim 1$ and are capable of setting GW limits, such as dielectric haloscopes \cite{millar_dielectric_2017, domcke_dielectric_2024} and dish antennas \cite{horns_searching_2013, bread, capdevilla2025highfrequencygravitationalwavesbread}.

We apply this extended framework to resonant cavity experiments, demonstrating spectral response calculations in the high and low frequency limits \TK{and investigate the accuracy of our approximate transformations for relevant classes of GW waveforms}. 
Finally, we discuss some limitations of our approach.

We use a natural unit system with $c=\epsilon_0=G=1$. Latin indices $i,j,\dots$ denote spatial components from $1$ to $3$, Greek indices $\mu,\nu,\dots$ denote space-time components from $0$ to $3$.
%
%
\section{Transformation to the Proper Detector Frame}
GW signals from distant astrophysical origins reach Earth as nearly ideal plane waves. Their waveform expressed in TT coordinates thus depends only on the retarded time $t_\text{ret}=t-z$, where $z$ denotes the distance from the coordinate origin along the line of sight to the GW source. For GWs of this form, the general transformation from TT coordinates into the PD frame is then, in linearized general relativity, given by \cite{Fortini1982}
\begin{align}\label{eq:TT_PD_transformation}
    h_{00}^\text{PD}&=-2\sum_{n=0}^\infty\frac{1}{(n+2)!}x^kx^lz^n\left[\partial_{z'}^nR_{0k0l}^\text{TT}\right]_{z'=0}\,,\nonumber\\
    h_{i0}^\text{PD}&=-2\sum_{n=0}^\infty\frac{n+2}{(n+3)!}x^kx^lz^n\left[\partial_{z'}^nR_{0kil}^\text{TT}\right]_{z'=0}\,,\\
    h_{ij}^\text{PD}&=-2\sum_{n=0}^\infty\frac{n+1}{(n+3)!}x^kx^lz^n\left[\partial_{z'}^nR_{ikjl}^\text{TT}\right]_{z'=0}\,,\nonumber
\end{align}
where the infinite sums arise from a Taylor expansion around $z\approx0$.
Here, we neglect accelerations or rotations of the PD frame, for which the transformation has been generalized in \cite{marzlin_fermi_1994}, assuming that these effects are much slower compared to $\omega_g$. 
For monochromatic GWs, the infinite sums in equations \eqref{eq:TT_PD_transformation} can be resummed into exponential functions.  In contrast, more complex, time-dependent GW signals generally do not permit such simplifications. However, if the only coordinate dependence of the waveform $h^\text{TT}_{\mu\nu}$ is the retarded time $h^\text{TT}_{\mu\nu}=h^\text{TT}_{\mu\nu}(t_\text{ret})$, the initial Taylor expansion can be reversed by completing the sum with additional terms. This yields a closed-form expression, given by 
\begin{widetext}
\begin{align}\label{eq:PD_general_chirp_strain}
h_{00}^\text{PD}&=x^kx^l\left[\frac{2}{z^2}R_{0k0l}^{(-2)}(z')\Big|_{z'=0}+\frac{2}{z}R_{0k0l}^{(-1)}(z')\Big|_{z'=0}-\frac{2}{z^2}R_{0k0l}^{(-2)}(z)\right]\,,\nonumber\\
h_{0i}^\text{PD}&=x^kx^l\left[\frac{1}{z}R_{0kil}^{(-1)}(z')\Big|_{z'=0}-\frac{2}{z^3}R_{0kil}^{(-3)}(z')\Big|_{z'=0}-\frac{2}{z^2}R_{0kil}^{(-2)}(z)+\frac{2}{z^3}R_{0kil}^{(-3)}(z)\right]\,,\\
h_{ij}^\text{PD}&=x^kx^l\left[-\frac{2}{z^2}R_{ikjl}^{(-2)}(z')\Big|_{z'=0}-\frac{4}{z^3}R_{ikjl}^{(-3)}(z')\Big|_{z'=0}-\frac{2}{z^2}R_{ikjl}^{(-2)}(z)+\frac{4}{z^3}R_{ikjl}^{(-3)}(z)\right]\,,\nonumber
\end{align}
\end{widetext}
where the notation $R^{(-n)}(z)=\int_{z_0}^zdz_1\dots\int_{z_0}^{z_{n-1}}dz_n\,R(z_n)$ is used and we choose without loss of generality $z_0=0$.
In this form, the linearized space-time curvature can be transformed into the local reference frame of an inertial observer. 
The linearized Riemann tensor components
\begin{equation}\label{eq:Riemann_lin}
R_{\mu\nu\rho\sigma}=\frac12(h_{\mu\sigma,\nu\rho}+h_{\nu\rho,\mu\sigma}-h_{\nu\sigma,\mu\rho}-h_{\mu\rho,\nu\sigma})\,,
\end{equation} 
can now be used to express the transformation in terms of the strain metric. 
Due to the invariance of the Riemann tensor components under coordinate transformations up to linear order in $h_{\mu\nu}$, we can use TT coordinates to express the strain metric. The non-zero components of $R_{\mu\nu\rho\sigma}$ for GWs are then given by 
\begingroup
\allowdisplaybreaks
\begin{align}\label{eq:riemann_components}
R_{0k0l}^\text{TT}&=-\frac12\partial_z^2h_{kl}^\text{TT}\,,\nonumber\\
R_{0kjl}^\text{TT}&=-\frac12\partial_z^2\left(h_{kj}^\text{TT}\delta_l^z-h_{kl}^\text{TT}\delta_j^z\right)\,,\\
R_{ikjl}^\text{TT}&=
\begin{aligned}[t]
    -\frac12\partial_z^2\big(&h_{ij}^\text{TT}\delta_l^z\delta_k^z+h_{kl}^\text{TT}\delta_i^z\delta_j^z\\ &-h_{il}^\text{TT}\delta_j^z\delta_k^z-h_{kj}^\text{TT}\delta_i^z\delta_l^z\big)\,,\nonumber
\end{aligned}
\end{align}
\endgroup
where the coordinate dependence of the metric is implicit. 
Inserting these components into the general form of the transformation in equations \eqref{eq:PD_general_chirp_strain} yields the metric components of arbitrary GWs in the PD frame
\begin{widetext}
\begin{align}\label{eq:exact_PD_components_from_TT}
h_{00}^\text{PD}&=\frac{x^kx^l}{z^2}\left[h_{kl}^\text{TT}(t-z)-h_{kl}^\text{TT}(t)+z\,\dot{h}_{kl}^\text{TT}(t)\right] \,,\nonumber\\
    h_{i0}^\text{PD}&=\frac{x^mz\,\delta_i^n-x^mx^n\delta_i^z}{z^2}\left[h_{mn}^\text{TT}(t-z)+\frac{1}{z}\int_t^{t-z}h^\text{TT}_{mn}(t')dt'+\frac{z}{2}\dot{h}_{mn}^\text{TT}(t)\right]\,,\\
    h_{ij}^\text{PD}&=\frac{\delta_i^m\delta_j^nz^2+x^mx^n\delta_i^z\delta_j^z-x^mz\,\big(\delta_i^n\delta_j^z+\delta_i^z\delta_j^n\big)}{z^2}\left[h_{mn}^\text{TT}(t-z)+\frac{2}{z}\int_t^{t-z}h^\text{TT}_{mn}(t')dt'+h_{mn}^\text{TT}(t)\right]\,,\nonumber
\end{align}
\end{widetext}
where we used the standard notation $\partial_t h_{ij}^\text{TT} = \dot{h}_{ij}^\text{TT}$. 
Equations \eqref{eq:exact_PD_components_from_TT} can be used to analytically transform general GW signals into the PD frame as long as there is a closed-form expression for their integral. For the special case of monochromatic GWs, the form derived in \cite{berlin_detecting_2023} is recovered. In cases where equations \eqref{eq:exact_PD_components_from_TT} cannot be evaluated analytically, they provide a more efficient form for numerical computation than equations \eqref{eq:TT_PD_transformation}, because the infinite series converges slowly when $\omega_gL_\text{det}\gtrsim1$. 
In the following section, we apply equations \eqref{eq:exact_PD_components_from_TT} to GW chirps from black hole coalescence.

%
%
\section{Application to compact binary mergers}\label{sec:compact_binaries}
Any binary system of compact objects with masses $m_1$ and $m_2$ emits GWs with a frequency related to its orbital frequency. As the gravitational radiation carries away energy, the two objects spiral towards each other, which increases the GW amplitude and frequency until the objects eventually collide. The larger the two objects are, the earlier they will collide, and the lower the maximally emitted GW frequency. For our analysis, we consider gravitational wave chirp signals of the form
\begin{equation}\label{eq:chirpStrainTT}
    h^\text{TT}_{ij}(t_\text{ret}) = h_0 (t_\text{ret})\,\hat{p}_{ij} \,e^{-i\left(\phi (t_\text{ret})+\phi_0 \right)}\,,
\end{equation}
where $t_\text{ret}=t-z$ is the retarded time before coalescence, $h_0 (t)$ is the amplitude of the GW strain, the matrix $\hat{p}_{ij} $ describes the GW polarization, $\phi (t)$ is the time evolving phase and $\phi_0 $ a constant relative phase. Equation \eqref{eq:chirpStrainTT} can also be extended to sums of multiple similar expressions. Note that only the real part of equation \eqref{eq:chirpStrainTT} is physical, but we are using complex notation for the metric in TT coordinates and take the real part after transforming into the PD frame. 

Upon the formation of a binary system, the initial orbit can in general be arbitrarily eccentric. However, GW emission leads to rapid circularization of the orbit, with \TK{the eccentricity scaling as $e\sim (a(t)/a_0)^{19/12}$ \cite{maggiore_gravitational_2008}, where $a$ is the semi-major axis decreasing with time and $a_0$ is the initial semi-major axis at binary formation.} 
Given that high-frequency GWs with $\omega_gL_\text{det}\gtrsim 1$ arise predominantly in the late inspiral phase, waveform models assuming circular orbits are of primary interest for detection.

In this work, we will thus focus on GWs created by binaries on quasi-circular Newtonian orbits \cite{maggiore_gravitational_2008}, where $h_0(t)=d^{-1}(5M_c^5/t)^{1/4}$ is the strain amplitude for the binary with chirp mass $M_c = (m_1m_2)^{3/5}/(m_1+m_2)^{1/5}$ at a distance $d$ and $\phi(t)=2(t/5M_c)^{5/8}$ is the phase dependence. For the GW polarization matrix we use
\begin{equation}
    \hat{p}_{ij}=\begin{pmatrix}
        (1+\cos^2\iota)/2&i\cos\iota&0\\
        i\cos\iota&-(1+\cos^2\iota)/2&0\\
        0&0&0
    \end{pmatrix}\,,
\end{equation} 
where $\iota$ is the angle between the line of sight and the angular momentum vector of the binary.
For GW signals of this type, equations \eqref{eq:exact_PD_components_from_TT} can be analytically evaluated using the integral
\begin{equation}
    \int_t^{t_\text{ret}}\frac{e^{-iBt^{b}}}{t^{a}}dt=\frac{\Gamma
   \left({\textstyle\frac{1-a}{b}},i Bt^b\right)-\Gamma
   \left({\textstyle\frac{1-a}{b}},i Bt_\text{ret}^b\right)}{(iB)^{\frac{1-a}{b}}b}\,,
\end{equation}
where $\Gamma(x_1,x_2)$ is the (upper) incomplete gamma function.

At the end of the inspiral phase, when the binary system approaches the innermost stable circular orbit (ISCO), perturbative solutions of Einstein’s equations are not sufficient and numerical GR must be used to model the signal. Thus, we restrict our calculation to times $t\gg t_\text{ISCO}$ where $\dot{\phi}(t)=\omega_g(t)\ll\omega_\text{ISCO}$. The frequency $\omega_\text{ISCO}$ depends on the compact objects forming the binary system and is approximately given by $\omega_\text{ISCO}\simeq (6\sqrt{6}\,(m_1+m_2))^{-1}$ for two black holes. For the case where $m=m_1=m_2$,
\begin{equation}\label{eq:t_ISCO}
    t_\text{ISCO}\simeq6.3\,\text{ms}\,\frac{m}{M_\odot}\,.
\end{equation} 

Although this model encapsulates the central features of chirp-like transient signals like a time-dependent phase and amplitude, the sensitivity to real signals with e.g. remaining eccentricity or orbital precession can be significantly enhanced by using more precise models \cite{Blanchet2024}. While the exact transformation to the PD frame is possible only for simple waveforms, we will show in the next section how to derive accurate approximations for different GWs of the form in equation \eqref{eq:chirpStrainTT}. 

%
%
\section{Approximate transformation}\label{sec:approximations}
Although equations \eqref{eq:exact_PD_components_from_TT} are valid for all plane gravitational waveforms in linearized GR, it is not always possible to express the integral of the signal using standard functions, as demonstrated in the example of GW chirps discussed previously. We will therefore discuss two approximations to the exact metric. First, we approximate the PD metric components by replacing the infinite sums in equations \eqref{eq:TT_PD_transformation} with exponential functions as in equation \eqref{eq:chirpStrainTT}, thereby avoiding integration of the signal. \TK{Second, we consider a monochromatic approximation where phase and amplitude are assumed to be constant.} 

In order to sum the expressions in equations \eqref{eq:TT_PD_transformation}, we use the Taylor expansion of the retarded time $t_\text{ret}$ for $z/t\ll 1$.\footnote{Since the exponential term in equation \eqref{eq:chirpStrainTT} oscillates rapidly in comparison to the amplitude $\dot{\phi}\gg|\dot{h}/h|$ for most compact binaries, it is tempting to employ a stationary-phase approximation instead where $\partial^n_zh_{ij}^\text{TT}\approx(-i\dot{\phi})^nh_{ij}^\text{TT}$ in equation \eqref{eq:TT_PD_transformation}. However, when $n\to\infty$, terms containing derivatives of $h_0$ dominate and equations \eqref{eq:TT_PD_transformation} cannot easily be resummed.} This linear approximation is well justified for a detector with a characteristic length $L_\text{det}\sim\mathcal{O}(1\,\text{m})$ as long as $t\gg 3.3\,\text{ns}\,L_\text{det}/(1\,\text{m})$ before coalescence. Furthermore, as seen in equation \eqref{eq:t_ISCO}, the merger phase, which requires numerical GR, is typically entered at earlier times, making it more restrictive than the $t_\text{ret}$ expansion. 

We separately expand both the phase of the exponential and the amplitude of the Riemann tensor components 
\begin{align}
 \partial_z^2h_{ij}^\text{TT}(t_\text{ret})&=H(t_\text{ret})\,\hat{p}_{ij}\,h_0(t_\text{ret})e^{-i\phi(t_\text{ret})}\\
 &\approx h_0(t)(H(t)-K(t)z)\,\hat{p}_{ij}\,e^{-i(\phi(t)-\omega_g(t)z)}\,,\nonumber
\end{align}
where we have introduced the time-dependent frequency of the GW $\omega_g(t)=\dot{\phi}(t)$ and the functions
\begingroup
\allowdisplaybreaks
\begin{align}\label{eq:general_trafo_eqs}
    H(t)&=\frac{\ddot{h}_0-2i \dot{h}_0 \dot{\phi}}{h_0}-\dot{\phi}^2-i\ddot{\phi}\,,\\
    K(t)&=\frac{\dddot{h}_0-2i\dot{\phi} \ddot{h}_0 -\dot{h}_0 \big(\dot{\phi}^2+3i\ddot{\phi}\big)}{h_0}-2\dot{\phi}\ddot{\phi}-i\dddot{\phi}\,,
\end{align}
\endgroup
where $K(t)=h_0(t)^{-1}\partial_t(h_0(t)H(t))$. In this form, the $z$-dependence simplifies such that the $n$-th derivative of the Riemann components can be evaluated analytically and summed up to condense the sums. For the simple chirp signals discussed in section \ref{sec:compact_binaries}, these functions become 
\begin{align}
    H(t)&=\frac{5}{(4t)^2}\left(1+\frac74i\phi(t)-\frac54\phi(t)^2\right)\,,\\
    K(t)&=-\frac{45}{64t^3}\left(1+\frac{91}{72}i\phi(t)-\frac{5}{9}\phi(t)^2\right)\,.
\end{align} 

As a result, the infinite summations in the transformation in equations \eqref{eq:TT_PD_transformation} can be rewritten using the function 
\begin{equation}
    F(x)=\frac{ 1+ix-e^{ix}}{x^2}=\sum_{n=0}^\infty\frac{(ix)^n}{(n+2)!}
\end{equation}
and its derivatives, such that the metric becomes 
\begin{widetext}
\begin{align}\label{eq:PD_chirp_strain}
    h_{00}^\text{PD}&=\Big[H(t)\,F\left(\omega_gz\right)
    +izK(t)\,F'\left(\omega_gz\right)\Big]\;x^kx^lh_{kl}^\text{TT}\,,\nonumber\\
    h_{i0}^\text{PD}&=\frac12\Big[H(t)\left(F\left(\omega_gz\right)-iF'\left(\omega_gz\right)\right)+izK(t)\left(F'\left(\omega_gz\right)-iF''\left(\omega_gz\right)\right)\Big]\Big[h_{ki}^\text{TT}zx^k-h_{kl}^\text{TT}x^kx^l\delta_i^z\Big]\,,\\
    h_{ij}^\text{PD}&=\Big[-iH(t)\,F'\left(\omega_gz\right)+zK(t)\,F''\left(\omega_gz\right)\Big]\Big[h_{ij}^\text{TT}z^2+h_{kl}^\text{TT}x^kx^l\delta_i^z\delta_j^z-h_{il}^\text{TT}\delta_j^z z x^l-h_{kj}^\text{TT}\delta_i^z z x^k\Big]\,.\nonumber
\end{align}
\end{widetext}
Again, only the real part of equations \eqref{eq:PD_chirp_strain} is physical. 
Furthermore, we have suppressed the time argument of $\omega_g=\omega_g(t)$ in equations \eqref{eq:PD_chirp_strain}, \eqref{eq:PD_chirp_strain_approx} for readability and the components $h^\text{TT}_{ij}=h_{ij}^\text{TT}(t)$ are implicitly evaluated at the spatial origin $z=0$.

Finally, we consider a further approximation to equations \eqref{eq:PD_chirp_strain} at large times before the merger when ${\phi(t)\gg1}$. In this limit, the functions in equations \eqref{eq:general_trafo_eqs} approach $H(t)\approx-\omega_g(t)^2$ and $K(t)\rightarrow0$. Then, the resulting components are given by
\begin{align}\label{eq:PD_chirp_strain_approx}
    h_{00}^\text{PD}&=-\omega_g^2\,F\left(\omega_gz\right)\;x^kx^lh_{kl}^\text{TT}\,,\\\nonumber
    h_{i0}^\text{PD}&=-\omega_g^2\frac{F\left(\omega_gz\right)-iF'\left(\omega_gz\right)}{2}\Big[h_{ki}^\text{TT}zx^k-h_{kl}^\text{TT}x^kx^l\delta_i^z\Big]\,,\\\nonumber
    h_{ij}^\text{PD}&=i\omega_g^2\,F'\left(\omega_gz\right)\\&\times\Big[h_{ij}^\text{TT}z^2+h_{kl}^\text{TT}x^kx^l\delta_i^z\delta_j^z-h_{il}^\text{TT}\delta_j^z z x^l-h_{kj}^\text{TT}\delta_i^z z x^k\Big]\,,\nonumber
\end{align}
and approximate equations \eqref{eq:PD_chirp_strain} to very high accuracy within the constraints $z\ll t$ and $t>t_\text{ISCO}$.
However, this expression is limited to the simple quasi-Newtonian binary model discussed previously. For example, as soon as waveforms for eccentric orbits and spinning compact objects with orbital precession are used, the amplitude $h_0(t)$ can be modulated on much shorter timescales \cite{Precession, Precession_2} and the full equations \eqref{eq:PD_chirp_strain} with equations \eqref{eq:general_trafo_eqs} must be used, which is demonstrated explicitly in section \ref{sec:validity}.

\TK{
Equations \eqref{eq:PD_chirp_strain_approx} are an example of a monochromatic approximation. It consists of expanding the phase as before $\phi(t_\text{ret})\approx\phi(t)+\dot{\phi}(t)(t_\text{ret}-t)$, and approximating 
\begin{equation}\nonumber
    h^\text{TT}(t_\text{ret})\approx h_0(t)e^{-i\phi(t)+i\dot{\phi}(t)z}=h^\text{TT}(t)e^{i\omega_g(t)z}\,.
\end{equation}
 For a slowly varying amplitude $\dot{h}_0(t)\ll h_0(t)\dot{\phi}(t)$ and a slowly accelerating phase $\ddot{\phi}(t)\ll\dot{\phi}(t)^2$, we can approximate all time derivatives $\partial_t\to-i\dot{\phi}$ and set $\dot{h}_0\approx0$ when calculating any derived quantities from $h^\text{TT}(t)$. 

Therefore, the PD metric components in equation \eqref{eq:PD_chirp_strain_approx} are related to those of monochromatic GWs as derived in \cite{berlin_detecting_2023} by replacing 
\begin{align}\label{eq:monochromatic_approximation}
    h_0e^{-i\omega_gt}&\to h_0(t)e^{i\phi(t)}\\
    \omega_gx^i&\to\dot{\phi}(t)x^i\nonumber
\end{align}   
in equations \eqref{eq:PD_chirp_strain}. 
We refer to the replacements in equation \eqref{eq:monochromatic_approximation}
 as a \emph{monochromatic approximation} in the rest of this work. It is a general method to calculate the transformation or quantities derived from the metric components, such as effective currents for a nearly monochromatic wave. As GW detectors are often designed and optimized assuming monochromatic GWs, their optimal performance extends to more complex waveforms only when the monochromatic approximation \eqref{eq:monochromatic_approximation} remains valid.} 

\begin{figure}
    \centering
    \includegraphics[width=0.4\textwidth]{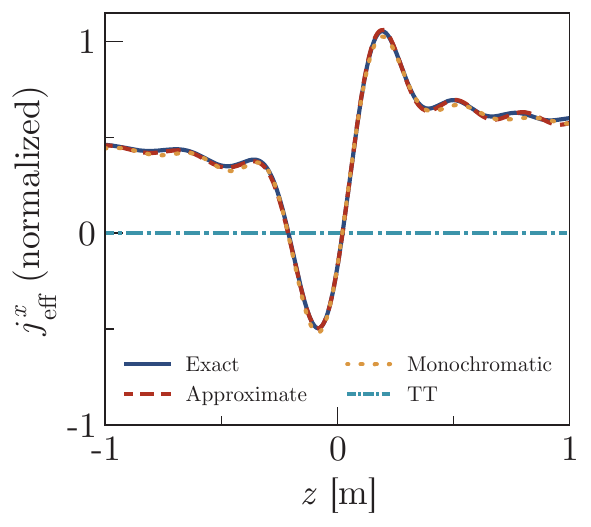}
   \caption{\textbf{Effective current of a chirp signal.} For a GW traveling along the $z$-axis with $M_c\approx10^{-6}M_\odot$, $\iota=\frac{\pi}{2}$, $\dot{\phi}(t)=\text{GHz}$ the effective current is calculated before coalescence in a cylindrical volume within a static magnetic field along the $z$-axis. The solid curve corresponds to $J_\text{eff}$ as obtained from equation \eqref{eq:exact_PD_components_from_TT}, the dashed curve to equation \eqref{eq:PD_chirp_strain}, the dotted curve to the monochromatic approximation \eqref{eq:monochromatic_approximation} and the dash-dotted curve to $J_\text{eff}^x$ evaluated in TT coordinates without approximation.}
  \label{fig:Jeff}
\end{figure}

\TK{As an example for an observable quantity derived from the GW metric components, we evaluate the component perpendicular to the direction of propagation of the GW for the effective electromagnetic current 
\begin{equation}\label{eq:Jeff}
    J_\text{eff}^\mu=(\rho_\text{eff},\,\bm{J}_\text{eff})^\mu = \partial_\nu\Big(\frac{h}{2}F^{\mu\nu}+h^\nu_{\,\rho} F^{\rho\mu}-h^\mu_{\,\rho} F^{\rho\nu}\Big)\,,
\end{equation}
which is induced by a GW passing through an electromagnetic field with field strength tensor $F^{\mu\nu}$. Certain detectors could observe this quantity and will be further discussed in chapter \ref{sec:sensitivities}.
In figure \ref{fig:Jeff} we compare the exact result with the approximation in \eqref{eq:PD_chirp_strain} and the monochromatic approximation \eqref{eq:monochromatic_approximation} for a specific chirp signal with chirp mass $M_c\approx10^{-6}M_\odot$. } 
Additionally, we show the effective current calculated for the signal in the TT frame.

In the limit of $z\rightarrow-\infty$ equations \eqref{eq:PD_chirp_strain_approx} reduce to a form that resembles the TT metric from equation \eqref{eq:chirpStrainTT} for $x,\,y\ll z$ and $x\,\omega_g(t_\text{ret}),\;y\,\omega_g(t_\text{ret})\ll 1$. This can be seen using equations \eqref{eq:TT_PD_transformation} and \eqref{eq:riemann_components} even without expanding $t_\text{ret}$ and by only keeping the lowest order terms in $1/z$. 
A similar behavior occurs for monochromatic waves at high frequencies. However, for chirp signals, $\omega_g(t_\text{ret})\rightarrow0$ as $z\rightarrow-\infty$. 
Nevertheless, provided that detectors can measure effects produced by gravitational waves near $z=0$, the TT limit remains invalid for sensitivity predictions.

%
%

\section{Range of Validity}\label{sec:validity}

\TK{The validity of the approximations in equations \eqref{eq:PD_chirp_strain} and the monochromatic approximation \eqref{eq:monochromatic_approximation} depend on the amount by which the phase and amplitude of the GW vary with $t_\text{ret}$. In order to understand this quantitatively, we introduce two representative examples which can be treated analytically. First, we consider a Gaussian burst, with $\phi(t)=\omega_0 t$ and $h_0(t)=e^{-(t/\tau)^2}$, where $\tau$ is the timescale of the burst. Second, we implement a quadratic phase modulation where $\phi(t)=\omega_0 t+(t/\tau)^2$ and $h_0(t)=h_0$. Both cases converge to a monochromatic wave as $\tau\to\infty$. However, similar waves with $\omega_0\sim\tau^{-1}$ can also contribute to physically expected waveforms from compact binaries in different scenarios \cite{chung-jukko_multimessenger_2024, Bezares:2018qwa, Dietrich:2018phi}. We measure the deviation of an approximate transformation to PD from the exact transformation \eqref{eq:exact_PD_components_from_TT} by the relative deviation of the resulting effective current within a volume V
\begin{equation}
    \Delta^2= \frac{1}{V}\int dV \frac{\lvert\bm{J}_\text{eff}^\text{exact}-\bm{J}_\text{eff}^\text{approx.}\rvert^2}{\lvert\bm{J}_\text{eff}^\text{exact}\rvert^2}\,.
\end{equation}

\begin{figure}
    \centering
    \includegraphics[width=0.4\textwidth]{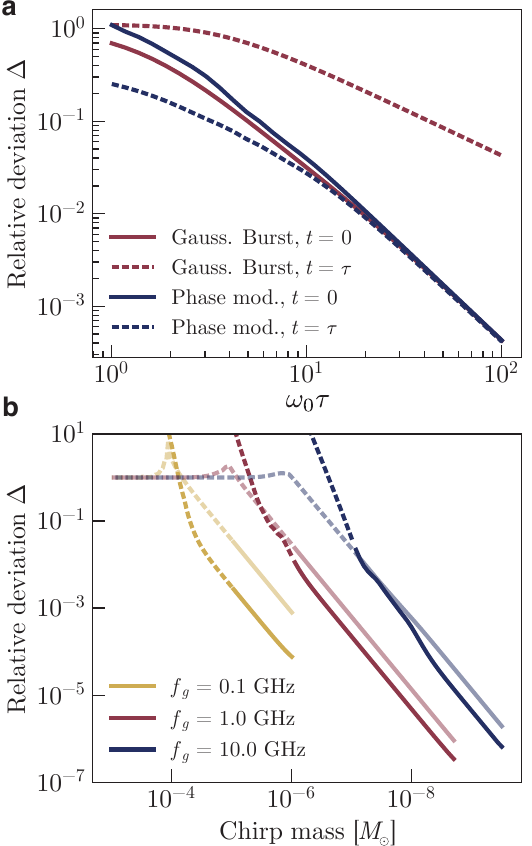}
   \caption{\textbf{Numerical analysis of the relative deviation of the effective current for the approximate metric components.} (a) The relative deviation of the effective current obtained for a gaussian burst and quadratic phase modulation for the monochromatic approximation \eqref{eq:monochromatic_approximation}. The GW is traveling along the same axis as a static magnetic field, for a length of $1\,\text{m}$  with $\omega_0=2\pi\,\text{GHz}$.
   (b) The relative deviation of the effective current obtained for a compact binary as described in section \ref{sec:compact_binaries}. The GW is traveling along the same axis as a static magnetic field, for a length of $1\,\text{m}$ and at different times to coalescence where $\dot{\phi}(t)=2\pi f_g$. The darker color corresponds to using equations \eqref{eq:PD_chirp_strain} and the lighter color to a monochromatic approximation \eqref{eq:monochromatic_approximation} for different frequencies $f_g$. The line is drawn dashed for unphysical combinations where $f_g>f_\text{ISCO}$.}
  \label{fig:ApproximationDeviation}
\end{figure}

The resulting deviation of a monochromatic approximation for a Gaussian burst and a quadratic phase modulation are shown in figure \ref{fig:ApproximationDeviation}a, where only the monochromatic approximation is considered for simplicity. The approximation remains valid if $\dot{h}_0\ll h_0\dot{\phi}$ and $\ddot{\phi}\ll\dot{\phi}^2$. For a Gaussian burst, this means $2t_\text{ret}/\tau\ll\omega_0\tau$ and for a quadratic phase modulation $2 t_\text{ret}/\tau+\omega_0\tau\gg\sqrt{2}$, as seen in figure \ref{fig:ApproximationDeviation}a. As expected, the monochromatic approximation is inaccurate as $\omega_0\tau\sim 1$ and converges to the exact result as $\tau\gg\omega_0$. This demonstrates that the exact transformation in \eqref{eq:exact_PD_components_from_TT} is only required when the GW's phase or amplitude is being modulated on a similar timescale as the carrier frequency. 

In the case of simple compact binaries, the monochromatic approximation is valid at times where $\phi(t_\text{ret})\gg1$ which is true for most of the high frequency GW parameter space as shown in figure \ref{fig:ApproximationDeviation}b. Furthermore, we find that the approximation in equation \eqref{eq:PD_chirp_strain} yields higher accuracy than the monochromatic approximation, as expected. Deviations $\Delta\gtrsim0.1$ only occur at unphysical times where $\dot{\phi}>\omega_\text{ISCO}$, shown as dashed lines. The monochromatic approximation also consistently exhibits larger deviations from the exact result than the approximation in equation \eqref{eq:PD_chirp_strain}.}

%
%
\section{Detector sensitivities to chirp signals
}\label{sec:sensitivities}

In this chapter, we apply the results from the previous sections to determine the spectral response of a large class of detectors to high-frequency GW signals. This requires the Fourier transforms of the GW signals. For chirp signals, where the phase typically changes more rapidly than the amplitude $|\dot{h}_0(t)/h_0(t)|\ll\dot{\phi}(t)\Leftrightarrow t\gg2\,\mu\text{s}\;m/M_\odot$, the Fourier transform can be evaluated with great precision by using a stationary phase approximation. The stationary point of the phase in the Fourier integral for $h_{\mu\nu}(f)$ is given at the time $t_\star(\omega)=\omega_g^{-1}(\omega)$. An expansion $\phi(t)$ to se\-cond order around $t_\star$ together with the approximation $h_0(t)\approx h_0(t_\star)$, yields the Fourier transform of equations \eqref{eq:exact_PD_components_from_TT} \cite{maggiore_gravitational_2008}
\begin{equation}\label{eq:h_fourier}
\begin{split}
    h_{\mu\nu}(\omega)=\sqrt{\frac{\pi}{2\ddot{\phi}(t_\star)}}|h_{\mu\nu}^{-}(t_\star)|\,e^{-i\left(\phi(t_\star)-\omega\,(t_\star+z)+\pi/4\right)}\,,
\end{split}
\end{equation}
where $h_{\mu\nu}^{-}$ are the components $\propto e^{-i\phi(t)}$ of equations \eqref{eq:exact_PD_components_from_TT}.

\begin{figure}[htbp]
  \centering
    \includegraphics[width=0.4\textwidth]{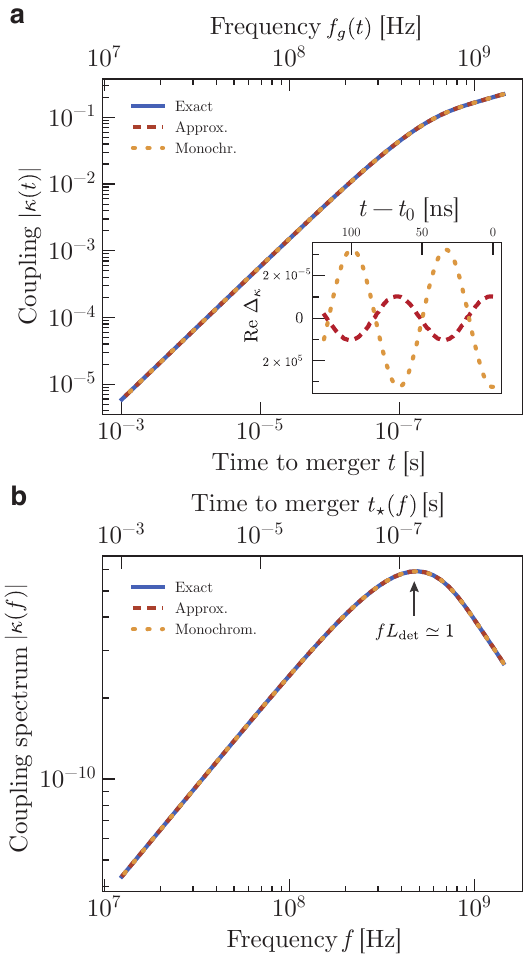}
    \caption{\textbf{Coupling coefficient of chirp signals to electromagnetic detectors.} We plot the coefficient $\kappa$ from equation \eqref{eq:coupling_coefficient} for a chirp signal with $M_c=10^{-6}M_{\odot}$ as shown in figure \ref{fig:Jeff}, traveling parallel to a static magnetic field aligned with the symmetry axis of a cylinder with length and radius of $1\,\text{m}$ with the resonant TE$_{212}$ mode. (a) We show the envelope of the time evolution of $|\kappa(t)|$ and the accuracy of the approximations according to equation \eqref{eq:Delta_kappa} close to the merger at $t_0$ in the inset. (b) The absolute value of the Fourier transform of $|\kappa(t)|$ in a stationary-phase approximation is shown.}
 \label{fig:kappa}
\end{figure}

As an example, we will focus on experiments that search for GW-induced perturbations of electromagnetic fields produced by the `inverse Gertsenshtein effect' \cite{gertsensthein}. 
In the absence of external charges and currents, the inhomogeneous Maxwell equations on a weakly perturbed space-time are modified and become 
\begin{align}\label{eq:GW_Maxwell_B}
    \nabla\cdot\bm{E}=\rho_\text{eff}\,,\\
    \nabla\times\bm{B}-\dot{\bm{E}}&=\bm{J}_\text{eff}\,,\label{eq:GW_Maxwell_E}
\end{align}
where the effective current in equation \eqref{eq:Jeff} is not coordinate invariant like $R_{\mu\nu\rho\sigma}$ and needs to be evaluated in the same frame as the $\bm{E}$- and $\bm{B}$-fields. Since $J_\text{eff}^\mu$ depends on the electromagnetic fields through the field strength tensor $F^{\mu\nu}$, the full equations are in general not simply solvable. However, most experiments set up strong background fields $F_0^{\mu\nu}$ which get $\mathcal{O}(h)$ corrections $F^{\mu\nu}=F_0^{\mu\nu}+F_\text{sig}^{\mu\nu}$ due to the GW. Therefore, to linear order in $h$, equations \eqref{eq:GW_Maxwell_B} and \eqref{eq:GW_Maxwell_E} can be decoupled and $J_\text{eff}^\mu$ evaluated solely in terms of the $F_0^{\mu\nu}$ fields. For example, when $\bm{E}_\text{sig}$ is a solenoidal resonant eigenmode of a cavity with $\nabla\cdot\bm{E}_\text{sig}=0$,  $\nabla^2\bm{E}_\text{sig}=-\omega_\text{sig}^2\bm{E}_\text{sig}$ and $\bm{E}_\text{sig}(\bm{x},t)=e(t)\bm{E}_\text{sig}(\bm{x})$, the signal amplitude obeys 
\begin{equation}\label{eq:GW_EM_EOM}
    \ddot{e}(t)+\omega_\text{sig}^2e(t)=-\frac{\int dV\bm{E}^*_\text{sig}\cdot\dot{\bm{J}}_\text{eff}}{\int dV |\bm{E}_\text{sig}(\bm{x})|^2}\,.
\end{equation}
The integral on the right-hand side of equation \eqref{eq:GW_EM_EOM} describes the geometric overlap of the GW induced current with the signal field, and is a generic feature of electromagnetic GW detectors. Even open and non-resonant systems as discussed in \cite{bread, millar_dielectric_2017} can conveniently be described with the same overlap integrals, where instead of a resonant mode, the reflected field distribution injected from the readout system is used for $\bm{E}_\text{sig}$ \cite{egge_axion_2023}. It is useful to define a dimensionless coupling coefficient
\begin{equation}\label{eq:coupling_coefficient}
    \kappa(t)=\frac{1}{f_g(t)^2\sqrt{V}}\frac{\int dV\bm{E}^*_\text{sig}(\bm{x})\cdot\dot{\bm{J}}_\text{eff}(t,\bm{x})}{h_0(t)B_0\sqrt{\int dV|\bm{E}_\text{sig}(\bm{x})|^2}}\,,
\end{equation}
for which the Fourier transform can be obtained with equation \eqref{eq:h_fourier}.
An example for such a coefficient is shown in figure \ref{fig:kappa}a. \TK{As expected from the results shown in figure \ref{fig:ApproximationDeviation}b, there is no notable deviation of the coupling coefficient when considering the two approximations we have discussed within the physical parameter space of quasi-Newtonian compact binaries. In the inset of figure \ref{fig:kappa}a, we show the remaining deviation from the exact coupling coefficient according to
 \begin{equation}\label{eq:Delta_kappa}
     \Delta_\kappa(t)=\frac{|\kappa_\text{exact}(t)-\kappa_\text{approx.}(t)|}{|\kappa_\text{exact}(t)|}\,.
 \end{equation}
 }\TK{We find a small relative deviation of the approximations to the exact result. Therefore, detectors which are optimized for the geometry of monochromatic waves, are usually simultaneously optimized for chirp signals as well. The overlap using the TT frame is not shown since $J_\text{eff}^\text{TT}=0$ in this case.}

Since PD coordinates are expanded around locally flat space, the metric perturbations in equation \eqref{eq:TT_PD_transformation} depend on the curvature components \eqref{eq:Riemann_lin} and decline at low frequencies $h^\text{PD}\propto\omega_g^2h^\text{TT}$ as seen in figure \ref{fig:kappa}b. 
Furthermore, the slope of $\kappa(f)$ at high frequencies arises since the chirp spends decreasingly less time oscillating at individual higher frequencies. Both effects balance around the frequency $\omega_gz\simeq 1$, leading to a maximum in the spectrum $\kappa(f)$. %
In resonant cavity experiments, eigenmode frequencies are usually of the same order of magnitude, allowing the most relevant frequency space to be probed. However, the broadband nature of the signal disfavors resonant experiments which are sensitive only to a narrow fraction of the GW spectrum \cite{aggarwal_challenges_2025}.

%
%
\section{Conclusions}
We have presented a general analytic form for the transformation of high-frequency gravitational waveforms, especially chirp-like signals, into the PD frame. This transformation is essential to model a detector response accurately in regimes where the GW wavelength becomes comparable to the size of the detector, and where the long-wavelength approximation can no longer be employed. Such considerations will be relevant for next-generation GW experiments targeting the MHz–GHz range.

We expressed the exact transformation from the TT frame into the PD frame in terms of the TT metric, its integral, and its derivatives (equations \eqref{eq:exact_PD_components_from_TT}). For cases where the integral cannot be computed analytically or where high accuracy is not required, we provided an alternative formulation based on a Taylor expansion of the retarded time until coalescence for small detector sizes (equations \eqref{eq:PD_chirp_strain}). This approximation eliminates the need for explicit integration, while it remains valid across a wide physical parameter space.

For quasi-Newtonian circular orbits long before merger, further approximations are typically applicable, yielding more simplified expressions. We validated these approximations by computing the spectral signal response in general electromagnetic detectors and demonstrated that quasi-monochromatic approaches preserve the geometric structure of the effective current, \TK{and approximate the transformation of quasi-Newtonian chirp signals well. However, signals whose phase or amplitude is modulated on a similar timescale as its oscillation, cannot be accurately approximated by a monochromatic wave and require the exact transformation we have derived.} 
Overall, the formulas given in equations \eqref{eq:exact_PD_components_from_TT} and \eqref{eq:PD_chirp_strain} provide practical tools for calculating the response of high-frequency gravitational wave detectors to general, time-dependent waveforms.

Although a numerical evaluation remains necessary in specific parameter regimes where integration and expansion techniques cannot be used, the derived metric form is more computationally practical than the general series representation.

Together, these results enable a comprehensive analysis and design of high-frequency GW detectors and improve the evaluation of their sensitivity to realistic astrophysical signals.

\subsection*{Acknowledgments}
The authors acknowledge support by the Deutsche Forschungsgemeinschaft (DFG, German Research Foundation) under Germany's Excellence Strategy - EXC 2121 `Quantum Universe' - 390833306.

\bibliography{references.bib}

\setcounter{figure}{0} 
\setcounter{equation}{0}

\renewcommand\thefigure{A\arabic{figure}} 
\renewcommand\thetable{A\arabic{table}}

\end{document}